\documentclass[conference]{IEEEtran}
\pdfoutput=1 
\usepackage{cite}
\usepackage{graphicx}
\usepackage{subcaption}
\usepackage{tikz}
\usepackage{cite}
\usepackage[T1]{fontenc} 
\usepackage[fleqn]{amsmath}
\usepackage[cmintegrals]{newtxmath}
\usepackage{bm}
\usepackage{mathtools}
\usepackage{color, colortbl}
\definecolor{Gray}{gray}{0.9}
\usepackage{makecell}
\usepackage{tabularx,ragged2e,booktabs}
\usepackage{pgfplots,pgfplotstable}
\usepackage[printwatermark]{xwatermark}
\usepackage{xcolor}
\usepackage{datetime}
\usepgfplotslibrary{statistics}

\pgfplotsset{width=8cm,height=6.5cm,compat=1.13,every axis/.append style={
			label style={font=\small},
			tick label style={font=\small},
            mark size=1pt
	}}
\newcommand{\squeezeup}{\vspace{-4.5mm}}
\IEEEoverridecommandlockouts
\IEEEpubid{\makebox[\columnwidth]{Version: March 2019\hfill} \hspace{\columnsep}\makebox[\columnwidth]{}} 
\begin{document}
\title{Towards Secure Slicing: Using Slice Isolation to Mitigate DDoS Attacks on 5G Core Network Slices
}
\author{\IEEEauthorblockN{Danish Sattar}
	\IEEEauthorblockA{\textit{Department of Systems and Computer Engineering} \\
		\textit{Carleton University} Ottawa, Canada \\
		danish.sattar@carleton.ca}
	\and
	\IEEEauthorblockN{Ashraf Matrawy}
	\IEEEauthorblockA{\textit{School of Information Technology} \\
		\textit{Carleton University} Ottawa, Canada\\
		ashraf.matrawy@carleton.ca}
}

\maketitle
\begin{abstract}
In this paper, we propose a solution to proactively mitigate Distributed Denial-of-Service attacks in 5G core network slicing using slice isolation. Network slicing is one of the key technologies that allow 5G networks to offer dedicated resources to different industries (services). However, a Distributed Denial-of-Service attack could severely impact the performance and availability of the slices as they could share the same physical resources in a multi-tenant virtualized networking infrastructure. Slice isolation is an essential requirement for 5G network slicing.

In this paper, we use network isolation to tackle the challenging problem of Distributed Denial-of-Service attacks in 5G network slicing. We propose the use of a mathematical model that can provide on-demand slice isolation as well as guarantee end-to-end delay for 5G core network slices.  We evaluate the proposed work with a mix of simulation and experimental work. Our results show that the proposed isolation could mitigate Distributed Denial-of-Service attacks as well as increase the availability of the slices. We believe this work will encourage further research in securing 5G network slicing.

\end{abstract}
\IEEEpeerreviewmaketitle
\section{Introduction}
\label{sec:intro}
\begin{figure}[!ht]
	\centering
	\includegraphics[width=\linewidth,keepaspectratio=true]{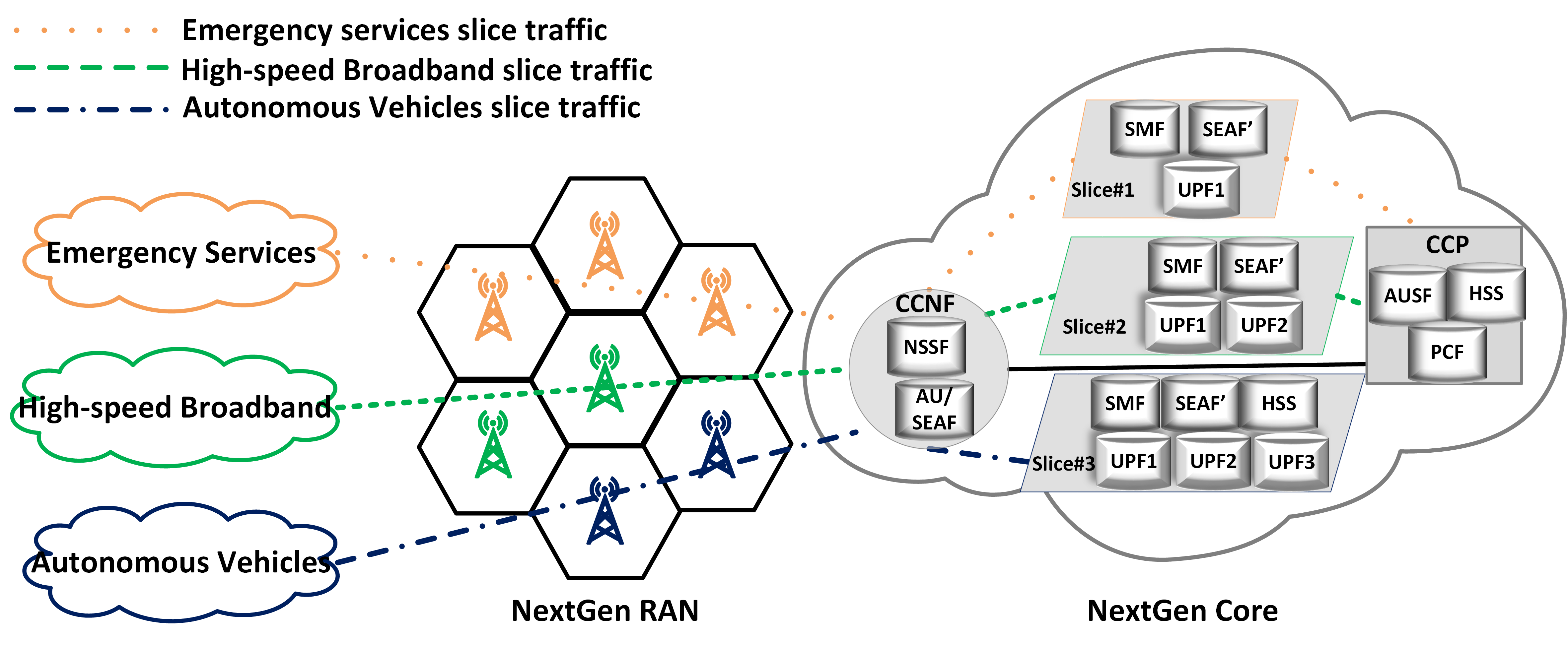}
	\caption{Logical view of network slicing in 5G Networks. NextGen Core is virtualized and the VNFs of slice\#1, slice\#2 and slice\#3 can be hosted on one or more physical servers. Elements used are: Network Slicing Selection Function (NSSF), Authentication Server Function (AUSF), Policy Control Function (PCF), Session Management Function (SMF), Authentication Unit (AU), Security Anchor Function (SEAF), Home Subscriber Server (HSS), User Plane Function (UPF) }
	\label{fig:slicingExample}
\end{figure}

The key motivation to present the work in this paper is that future mobile networks must provide services to a vast array of applications and devices with {\it competing} and perhaps {\it conflicting} requirements while simultaneously allowing flexible and agile deployment.  
5G networks deal with this issue in part by relying on {\bf network slicing} which has emerged as a key-enabler for proving heterogeneous services with different requirements to enable agile and scalable deployment of the 5G network (applications) \cite{8491249}. Network slicing takes advantage of virtualized infrastructures, where multiple services can be hosted on the same physical infrastructure, while at the same time guaranteeing service level agreements (SLAs) therefore allowing flexible and efficient utilization of limited resources.
In a 5G network, a slice could be instantiated for a specific service. Fig. \ref{fig:slicingExample} shows examples of different slices for autonomous vehicles (with a very low end-to-end delay requirement),  high-speed mobile broadband (with high bandwidth requirements), or a dedicated slice for emergency services. In Fig \ref{fig:slicingExample}, the Common Control Network Functions (CCNF) are shared among all three slices, and the Common Control Plane (CCP) is only providing services to slice\#1 and slice\#2.

A shift in the mobile network paradigm also mandates changes in the {\bf security} architecture with new security challenges that do not exist in the past or present mobile networks \cite{8039298}. In the current mobile network, if an attacker launches flooding Distributed Denial of Service (DDoS) attack against a mobile service, only that service will be affected, and there is a very low risk of host resource starvation-based attacks in current mobile networks (i.e., usually mobile network infrastructure is not virtualized). However, in a 5G network, a DDoS against one slice might affect other services because they could be tenants of the same shared virtualized infrastructure. An important requirement for 5G network slicing is {\bf isolation}~\cite{8039298}. Network slice isolation requires host resource isolation and network communication isolation that could be achieved using various techniques suggested in~\cite{8104638}. Each suggested technique has its advantages and disadvantages.  However, separating hardware resources provides strong isolation.  To the best of our knowledge, the work proposed in this paper is the first in 5G core network slicing that proposes and evaluates the use of slice resource isolation to mitigate DDoS attacks.

Distributed Denial-of-service attacks are difficult to mitigate. During a flooding DDoS attack, the attacker will try to overwhelm the target network or service by sending a large amount of traffic. The challenge is to keep the target network or service available to the end users. A few ways to mitigate flooding DDoS attacks are; the target network or service keep upscaling its resources, or telling upstream routers (ISPs) to block the traffic near the sources to reduce the impact. However, these solutions might not be possible in all scenarios. Some recent examples are DDoS attack on Github in 2015~\cite{Github2015} and 2018~\cite{Github2018}, and Dyn in 2016~\cite{Dyn2016}. The host resource starvation attack in a virtualized environment is another type of DoS attack, where attacker try to exhaust the hardware resources of the host \cite{8301700} that could result in an in-direct DoS attack on other hosted services (on the same host).

In 5G virtualized network, multiple tenants will share the same physical resources. A DDoS attack on one slice could disrupt the services hosted on other slices. There could be two attack vectors that can be used to affect other slices. If the communication link is shared between multiple slices, then a DDoS attack on one slice could impact other slices. The other attack vector would be if both slices are hosted on the same host then the attacker can take advantage of the co-hosting and launch an attack on other slices. For instance, increased Central Processing Unit (CPU) utilization of a slice could have an adverse effect on the performance and availability of other slices. It is possible to be hosted on the same host and not share the same communication link. In the worst case, multiple resource-intensive slices could be hosted on the same host and have a common communication link.

In this paper, we try to mitigate the Distributed Denial-of-Service (DDoS) attacks using slice isolation. We utilize a mathematical model to solve a security problem. In our proposed solution, we use slice isolation as security constraints for the optimization model and proactively mitigate DDoS attacks.  Our \textbf{contributions} are: (1) inter-slice isolation (complete separation of host hardware resources between slices), (2) intra-slice isolation (separation of host hardware resources between the slice components), (3) guaranteeing the end-to-end delay requirement for the slice and (4) optimal allocation of slices to efficiently utilize available hardware infrastructure.

Inter-slice isolation provides mitigation against the DDoS attacks because the hardware resources are not shared between the slices and DDoS on one slice does not impact the other slices. Whereas, intra-slice isolation provides better availability for the slices since the components of the slice are hosted on different hosts. Enabling intra-slice could reduce the impact of DDoS attack and increase the availability of the target slice as well as faster recovery of the service because the slice operator may not need to deploy reactive mitigation defense for all slices (e.g., migrating only affected slices).

The objective of our work is to provide inter-slice and intra-slice isolation, guarantee the end-to-end delay, make efficient use of system resources, and increase the availability of a slice against the DDoS attacks. To achieve these objectives, we propose the use of a security-focused optimization model that allocates slices and fulfills different requirements of the 5G network slicing. We use a combination of simulation and experimental work to evaluate our DDoS mitigation solution. We note that we do not use slicing in 5G Radio Access Network (RAN), we only focus on core slicing.

The rest of this paper is organized as follows. In Section~\ref{sec:relatedwork}, we present the literature review on 5G slicing security and the virtual network function placement in mobile networks. In section~\ref{sec:Tmodel} we discuss our threat model for the 5G network slice isolation. Section~\ref{sec:mathmodel} formulates optimization model for 5G network slicing. We discuss our evaluation methodology and experiment setup in section \ref{sec:EES}. In the section~\ref{sec:resutls}, we discuss our results and lastly, section~\ref{sec:con}, we present our conclusion.

\section{Related Work}
\label{sec:relatedwork}

{\bf Network Slicing Security: } V. Sathi \textit{et al.}~\cite{Sathi:2018:NPS:3242102.3242135} proposed a novel protocol for securing intra-slice network communication. The proposed protocol is based on modified proxy re-encryption and bilinear pairing on an elliptic curve. It provides mutual trust between participating entities as well as a distributed association between slice components in a secure manner (if central orchestrator is unavailable). Intelligent security focused auto-scaling algorithm has been proposed by Y. Khettab \textit{et al.}~\cite{8377298}. Authors leveraged the Network Function Virtualization (NFV) and Software-Defined Network (SDN) architecture to secure a network slice. SDN controller ONOS is used to enable security as a service in the proposed algorithm. The auto-scaling solution takes into consideration the startup time of the VM as well as guaranteeing the resources at the orchestrator level by setting minimum and maximum resources for each slice. IoT is one of the major drivers behind 5G network development. An efficient and secure method is required to authenticate a massive number of low powered devices. J. Ni \textit{et al.}~\cite{8314666} proposed a secure service-oriented authentication framework for 5G-enabled IoT devices ($ES^3A$). The proposed framework guarantees privacy-focused secure slice access and selection as well as a secure session key exchange mechanism between IoT devices. A quantitative analysis of network slice isolation has been proposed by Z. Kotulski \textit{et al.}~\cite{8511163}. The proposed framework is a layered structure, and each layer has its network elements and isolation level. Authors defined mathematical rules to calculate slice isolation.

{\bf Virtual Network Function Placement in Mobile Networks: } Recently, Virtual Network Function (VNF) placement in the virtual mobile network has gained research communities' interest. The main driving force is the architecture of the 5G network that heavily relies on virtualization. V. Sciancalepore et al. \cite{2018arXiv180103484S} have proposed a practical implementation of network slicing. In the proposed model, authors assume that in any given time slot, tenants can only request a single network slice. The proposed model is built on the multi-armed bandit problem, and the authors introduced three variations of the multi-armed bandit to solve various aspects of the network slice allocation. They simulated the optimization model using MATLAB and compared the results with a greedy algorithm. They also provided proof-of-concept implementation of the network slicing.
D. Dietrich \emph{et al.}~\cite{7945385} proposed a linear programming formulation for the placement of VNFs in the LTE core network. In the proposed algorithm, they provided a balance between optimality and time complexity. The authors R. Ford \emph{et al.}~\cite{DBLP:journals/corr/FordSMJR17} proposed an optimal VNF placement for the SDN-based 5G mobile-edge cloud. Their optimization algorithm provides resilience by placing VNFs in distributed data centers. They demonstrate in the evaluation that their algorithm can reduce 75\% of redundant data center capacity while still maintaining the same resilience.

\section{Threat Model}
\label{sec:Tmodel}
In this section, we describe the threat model used in this paper. Our threat model considers the 5G elements that are involved in network slicing.
\subsection{Assumptions}
Our threat model assumes that:
\begin{itemize}
	\item Network slicing is supported by the target network
	\item It is possible to modify the VNF configuration of a slice
	\item A network slice supports the installation of user-defined functions
	\item The 5G core network uses virtualized network functions
	\item The EPC supports migration of the slice components
	\item On the servers, multi-tenancy is supported by the slice operator
	\item It is possible to have slice-level authentication
	\item It is not possible to modify the hardware configurations of the baremetal systems
	\item The slice operators' network is functional
\end{itemize}
\subsection{Adversaries}
Our model considers two types of adversaries.
\label{sec:adv}
\begin{itemize}
	\item \textbf{Adversaries with administrative control:} An Adversary could be someone with administrative control of the slice. This type could be a tenant, an insider (assigned by the mobile operator to manage the slice), or an external attacker who has compromised a slice.
	\item \textbf{External adversaries:} External attacker could be a user of a slice or an external entity that could attack a slice.
\end{itemize}
\subsection{Attacks on a slice in the multi-tenant networking environment}
\label{sec:attacks}
In this threat model, we consider two of the possible attacks on a slice in the multi-tenant environment. These are two attacks we used in our experimental evaluation.
\subsubsection{DDoS flooding attack}
\label{sec:FloodDDoS}
\begin{figure}[!ht]
	\centering
	\includegraphics[width=\linewidth,keepaspectratio=true]{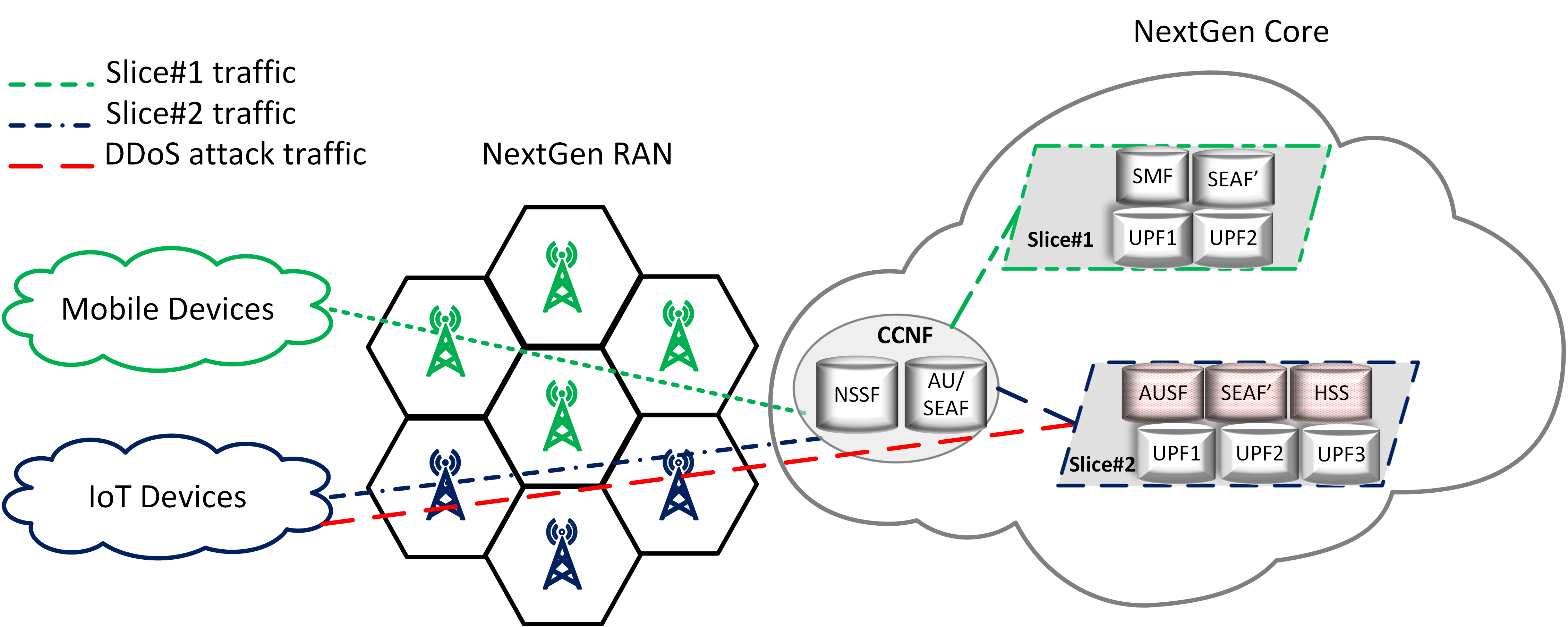}
	\caption{Logical view of DDoS attack on a slice} 
	\label{fig:smallcell1}
\end{figure}

Unlike LTE networks, which are usually single tenant, multi-tenancy in 5G networks will be a typical environment. For the flooding attacks, we assume that the attacker is an external adversary (second type). The attacker launches a DDoS attack that could flood the communication link(s) of the target slice (e.g., UDP flooding, TCP sync attack, or any other type of DDoS flooding attack). DDoS flooding attacks~\cite{7289347}, \cite{8301700} are not new or limited to the mobile networks. They have existed in traditional networks, cloud networks, mobile ad-hoc networks, software-defined networks and in many other types of networks. In 5G networks, network slicing presents a new attack vector to perform DDoS attacks.
A high-level view of the DDoS attack is shown in Fig.~\ref{fig:smallcell1}.

In Fig.~\ref{fig:smallcell1}, we have two slices providing services to two different class of devices (i.e., Mobile devices and IoT devices). The Common Control Network Functions (CCNFs) shared among both slices and provide slice selection and authentication functions. In this scenario, we have a large number of IoT devices attempting to authenticate with the slice\#2. This could result in a DDoS attack on slice\#2's components (i.e., AUSF, SEAF', and HSS). However, If both slices share the same physical link in the core network (all or some components of both slices could share the same physical link), then slice\#1 might be impacted, and it could result in an in-direct DDoS on the slice\#1.
\subsubsection{Slice-initiated attack}
\label{sec:sliceinitidated}
There is a possibility that a slice might initiate the attack in this multi-tenant environment. A slice-initiated attack would require administrative privilege because it might require modification of VNF/slice configuration. We assume that the attacker has administrative control of the slice (first type of adversary). An attacker could exhaust the resources of the VNF (e.g., 100\% CPU or RAM utilization)~\cite{8301700}, \cite{7289347}. The effect of a slice-initiated attack is limited to the slices hosted on the same host(s). If one or more components of a slice are hosted on the same physical host as the adversarial slice VNF(s), then sustained high utilization of hardware resources could degrade the performance of other slices.
\section{Proposed Solution}
\label{sec:propsol}
In this section, we describe our proposed solution, where we provide optimal slice allocation and slice isolation while fulfilling several requirements of 5G networks. We use intra-slice isolation to provide more reliability and better availability, inter-slice isolation for providing isolation between slices.

In the proposed solution, when slice allocation is requested, the slice could require intra-slice isolation for high availability. If there is a DDoS attack on the slice or a hardware failure, it could have catastrophic effect if the entire slice is allocated on one host. An essential requirement for 5G network slicing is isolation~\cite{3gpp.33.899}. The 3GPP states that "\emph{The 3GPP System shall have the capability to provide a level of isolation between network slices which confines a potential cyber-attack to a single network slice}" \cite{3gpp.33.891}. However, how network isolation should be achieved is left up to the implementation of the network slicing. In our solution, if a slice requires complete isolation from other slices, the inter-slice isolation constraint provides such functionality.

Currently, in our system, all the slice requests and isolation requirements are provided as input to the optimization solution. The optimization solution provides us with an allocation scheme for all the slices. We manually allocate all slices according to the allocation scheme. By using intra-slice and inter-slice constraints in our solution, we proactively try to mitigate DDoS attacks.

\section{Optimization Model Formulation}
\label{sec:mathmodel}
In this section, we explain the optimization model used in this paper.  Inter-slice isolation, guaranteeing end-to-end delay, and high reliability are essential requirements in 5G network slicing \cite{5gamericas}. Our security-focused optimization constraints fulfill different isolation requirements. Inter-slice isolation will depend on how the slicing is implemented. If multiple VNFs are assigned to the same VM from different slices, then one way to achieve inter-slicing isolation would be to assign the only slice per VM (container). In this paper, we consider one VNF per VM (container); therefore it already provides some level of isolation. However, we consider complete separation of hardware inter-slice isolation (such that no components of different slices can be allocated to the same host).

\subsection{Network and Request Model}
We use an undirected graph $G_p=(N_p,L_p)$ to represent the physical 5G core network topology. All the nodes in the network (i.e. servers, switches, routers and other devices present in the network) are represented by $N_p$, and $L_p$ denotes all the physical links between the nodes. A slice request is denoted by a directed graph $G_v = (N_v,L_v)$, where $N_v=(N_c\cup N_d)$ contains all the slice virtual network functions, the control and data plane virtual functions are represented by $N_c$ and $N_d$, respectively and $L_v$ represents requested links. Each edge in the directed graph is represented by $(i,j) \in L_v$. Each slice request is associated with end-to-end delay ($d_{E2E}$), intra-slice isolation ($K^c_{rel},K^d_{rel}$), inter-slice isolation ($\gamma^c,\gamma^d$) and each VNF in a slice is associated with a computing demand ($R^i$), and bandwidth (BW) requirement between VNF i and VNF j ($R^{ij}$). The description of all variable is provided in table~\ref{tab:parameters}.
\begin{table}[htbp]
	\centering
	\captionof{table}{Variable Description} \label{tab:parameters}
	\begin{tabular}{|r|l|}
		\hline
		Parameter & Description \\ \hline
		$N_p$ & Set of physical Nodes \\
		$L_p$ & Physical links between nodes \\
		$\sigma_k$ & Current CPU allocation of physical node $k$ \\
		$\sigma_{ef} $ & Curr. BW allocation of physical link between nodes $e,f$\\
		$\sigma_k^{max}$ & Maximum CPU capacity of physical node $k$ \\
		$\sigma^{max}_{ef} $ & Max. BW capacity of physical link between nodes $e,f$\\
		$T_{ef}$ & Physical link delay between node $e,f$ \\
		$\Delta_k$ & Physical node $k$ processing delay \\
		$\Delta^i$ & VNF $i$ processing delay \\
		$N_c$ & Requested set of slice control plane functions \\
		$N_d$ & Requested set of slice data plane functions \\
		$N_v$ & Requested set of slice VNFs ($N_c\cup N_d$) \\
		$L_v$ &  Requested virtual links of a slice\\
		$R^i$ & Requested CPU resource by a VNF $i$ \\
		$R^{ij}$ & Requested BW resource between VNF $i,j$\\
		$d_{E2E}$ & Requested End-to-End delay\\
		$K^c_{rel}$ & Requested intra-slice isolation for Control Plane\\
		$K^d_{rel}$ & Requested intra-slice isolation for Data Plane\\
		$\gamma^c$ & Requested inter-slice isolation for Control Plane \\
		$\gamma^d$ & Requested inter-slice isolation for Data Plane \\
		\hline
	\end{tabular} \par
	\bigskip
\end{table}
\subsection{Objective function}
In earlier work \cite{DBLP:journals/corr/abs-1802-04655}, we used a similar objective function and general constraints. They are influenced by prior work \cite{7945385} by Dietrich et al., especially the first term of the objective function and general constraints \ref{eq3}, \ref{eq4}, and \ref{eq6}.
\begin{equation}\label{eq1}
\begin{multlined}
\emph{Minimize}\\
\sum _{i\in{N}_{v}}\sum _{k\in {N}_{p}} \left(\sigma_k+ R^i\right)u^i_k
+\sum_{(i,j)\in{L}_{v}}\sum _{\substack{(e,f)\in {L}_{p} \\(e\neq f)}}T_{ef} y^{ij}_{ef}
\end{multlined}
\end{equation}

The objective function (\ref{eq1}) is to allocate the slice to least utilized physical nodes and find minimum delay path. The first term will assign the slice request to the least utilized servers. The second term will find the minimum delay path. Minimizing both terms will result in the assignment of a network slice to the least utilized servers, and it will find a path with least delay between the slice components.

The objective function is subjected to several Mixed-Integer Linear Programming (MILP) general and security-related constraints that we will explain next.
\subsection{General Constraints}
\begin{enumerate}
	\item \textbf{Slice Assignment, Placement and Resource Budget}
	\begin{equation}\label{eq2}
	\begin{multlined}
	\sum _{k \in N_p} u^i_k=1 \hspace{.5cm}\forall i \in N_v
	\end{multlined}
	\end{equation}
	\begin{equation}\label{eq3}
	\begin{multlined}
	\sum _{i \in N_v} \left(R^i + \sigma_k\right) u^i_k \leq \sigma^{max}_k \hspace{0.5cm}\forall k \in N_p
	\end{multlined}
	\end{equation}
	\begin{equation}\label{eq4}
	\begin{multlined}
	\sum _{(i,j) \in L_v} (R^{ij}+\sigma^{ij}_{ef}) y^{ij}_{ef} \leq \sigma^{max}_{ef} \hspace{0.5cm} \forall (e,f) \in L_p
	\end{multlined}
	\end{equation}
	\begin{equation}\label{eq5}
	\begin{multlined}
	\sum _{i \in N_v} R^i \leq \sum_{k \in N_p} \left(\sigma_k^{max} - \sigma_k\right)
	\end{multlined}
	\end{equation}
	\begin{equation}\label{eq6}
	\begin{multlined}
	\sum _{\substack{f\in {N}_{p} \\(e\neq f)}}\left(y^{ij}_{ef}-y^{ij}_{fe}\right)=\left( u^i_e-u^j_e\right)\\
	\hspace{0.5cm} i\neq j, \forall (i,j) \in L_v, \forall e \in N_p
	\end{multlined}
	\end{equation}
	
	\begin{equation}\label{eq7}
	\begin{multlined}
	u_k^i \in \{0,1\} \hspace{.5cm} \forall k \in N_p,  \forall i \in N_v
	\end{multlined}
	\end{equation}
	\begin{equation}\label{eq8}
	\begin{multlined}
	y_{ef}^{ij} \ge 0 \hspace{.5cm} \forall (i,j) \in L_v, \forall (e,f) \in L_p
	\end{multlined}
	\end{equation}
	The constraint (\ref{eq2}) ensures that each VNF is assigned to an exactly one server. Constraints (\ref{eq3}) and (\ref{eq4}) guarantee that allocated VNF resources do not exceed the physical servers' processing capacity and link bandwidth, respectively. A slice CPU demand should not exceed the remaining CPU capacity of the entire system. This is ensured by constraint (\ref{eq5}) since partial allocation of a slice is not the desired behavior. The conservation of flows, i.e., the sum of all incoming and outgoing traffic in the physical nodes that do not host VNFs should be zero is enforced by the constraint (\ref{eq6}), and this constrains also ensures that there is a path between VNFs. Constraints (\ref{eq7}) and (\ref{eq8}) ensures that $u_k^i$ and $y_{ef}^{ij}$ are binary and integer, respectively.
	\item \textbf{End-to-End Delay:}
	\begin{equation}\label{eq12}
	\begin{multlined}
	\left(\sum _{\substack{(i,j)\in {L}_{v} }} \sum _{\substack{(e,f)\in {L}_{p}\\e\neq f}}T_{ef}y^{ij}_{ef} +  \sum_{i\in N_v} \left(\Delta^i +\sum_{k\in N_p} \Delta_k u_k^i \right)\right)\leq d_ {E2E}
	\end{multlined}
	\end{equation}
	\begin{equation}\label{eq12a}
	\begin{multlined}
	T_{ef}=\frac{\sigma_{ef}}{\sigma^{max}_{ef}} \delta  + T_{ef,init} \hspace{.3cm} \forall (e,f) \in L_p
	\end{multlined}
	\end{equation}
	Constraint (\ref{eq12}) guarantees end-to-end delay for a slice in the current network state. End-to-end delay includes link delay, VNF processing delay, and physical node processing delay. Each time when a virtual link $(i,j)\in L_v$ is assigned to a physical link $(e,f)\in L_p$, it increases $T_{ef}$. $T_{ef}$ is a function of link utilization, and it is calculated using eq. (\ref{eq12a}), where $T_{ef,init}$ is the initial delay assigned to the link $(e,f)\in L_p$ and $\delta$ is the maximum increase in delay. We explain in section \ref{sec:evalmethod} how these values are calculated empirically.
\end{enumerate}
\subsection{Security-related Constraints}
\begin{enumerate}
	\item \textbf{Intra-Slice Isolation}
	\begin{subequations}\label{eq10}
		\begin{align}
		\sum _{i \in N_c} u^i_k\leq K_{rel}^c \hspace{.5cm}\forall k \in N_p, K_{rel}^c=1,2,3...\\
		\sum _{i \in N_d} u^i_k\leq K_{rel}^d \hspace{.5cm}\forall k \in N_p, K_{rel}^d=1,2,3...
		\end{align}
	\end{subequations}
	It might be required to have different levels of intra-slice isolation for control and data plane. In constraints (\ref{eq10}), $K_{rel}^c$ and $K_{rel}^d$  ensure the intra-slice isolation for control plane and data plane, respectively. Intra-slice isolation can improve the availability of a slice. For instance, if all components are hosted on one host, then a DDoS on one component could lead to complete unavailability of the slice as well as longer migration time (if required). However, if there was some level of isolation then DDoS on a single component might not result in complete unavailability and the operator would not have to migrate the entire slice but some components which could result in faster recovery.
	
	\item \textbf{Inter-Slice Isolation}
	\begin{subequations}\label{eq11}
		\begin{align}
		\hspace{-0.3cm} \sigma_k u^i_k=0\iff\gamma^c=1 \hspace{.2cm}\forall k \in N_p, \forall i \in N_v,\gamma^c \in \{0,1\}\\
		\hspace{-0.3cm} \sigma_k u^i_k=0\iff\gamma^d=1 \hspace{.2cm}\forall k \in N_p, \forall i \in N_v,\gamma^d \in \{0,1\}
		\end{align}
	\end{subequations}
	Inter-slice isolation could provide more security in a multi-tenant environment. Constraint (\ref{eq11}) provides the required hardware inter-slice isolation. In the allocation scheme, it might be required to have inter-slice isolation between control plane function (for security reasons) but not in the data plane functions or vice versa. Inter-slice isolation for control plane or data plane is only provided when $\gamma^c=1$ or $\gamma^d=1$, respectively. We note that our inter-slice isolation does not provide bandwidth isolation.
\end{enumerate}
\section{Evaluation and Experimental Setup}
\label{sec:EES}
The implementation of a complete 5G network is still in early stages and evaluating a fine-grained network slicing solution like ours in a 5G network is very challenging. There are very few testbeds and even fewer available to academia for testing and experimentation. We are not aware of any publicly available simulator or emulators that can provide complete simulation or emulation of 5G RAN and core network with slicing support. Another challenging issue is the cost of 5G hardware and software. Therefore, we made the best use of currently available tools (open source) and generic hardware to evaluate our solution.

\subsection{Experiment Setup}
\begin{figure*}[ht]
	\centering
	\includegraphics[width=15cm,height=4.8cm]{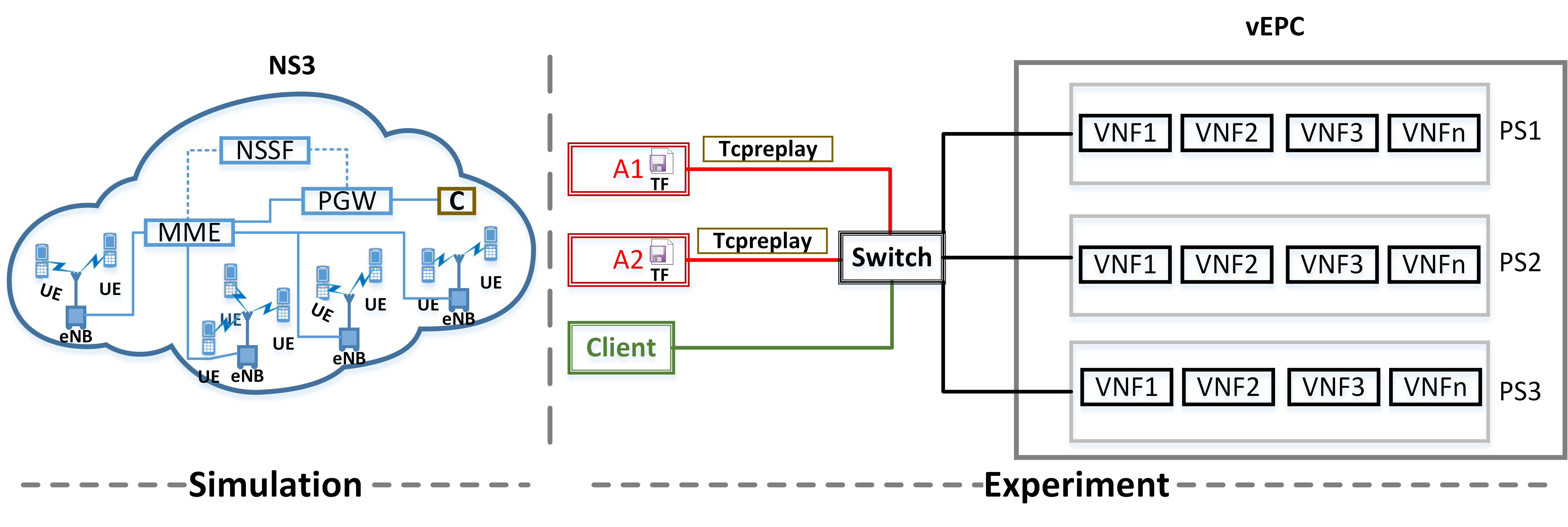}
	\caption{Evaluation Setup. The components used in the ns3 simulation are User Equipment (UE), evolved NodeBs (eNB), Mobility Management Entity (MME), Packet Data Gateway (PGW) and \textbf{C} is a traffic trace collection node. PS1-PS3 are the slice allocation servers, and A1-A2 are DDoS nodes. \textbf{TF} (pcap) is the traffic trace file collected in the simulation at \textbf{C} and replayed at A1 and A2 using \emph{tcpreplay}}
	\label{fig:topology}
\end{figure*}

To evaluate our solution, we developed a testbed using six physical servers. We used three servers to allocate slices, two as DDoS nodes and one as a client as shown in Fig. \ref{fig:topology}. The hardware specification for PS1, PS2, PS3, A1, and A2 are: Intel(R) Xeon(R) CPU E5420 @ 2.50GHz (4 cores), RAM 8GB, and network bandwidth 100Mb/s. In each of our experiments, we deploy 12 core network slices; each slice consisting of three components (network functions). The slice allocation is static in our experiments where we input all the slice requests and requirements in our optimization algorithm, and it provides us with an allocation scheme that we use for the entire experiment. AMPL \cite{AMPL} is used to model the optimization algorithm, and CPLEX (ilogcp) 12.8.0 is used as MILP solver. OpenVZ \cite{OpenVZ} is used for virtualization. It is an open source container-based virtualization platform. OpenVZ allows each container to have a specific amount of CPU, RAM, and Hard Drive (HDD). Each container (which hosts one VNF) performs and executes like a stand-alone server. We have installed the CentOS 6 \cite{CentOS} operating system in every container. We used Linux Traffic control (\emph{tc}) \cite{tc} to allocate bandwidth for each container.

\subsection{Combining simulations with the use of testbed}
As demonstrated in Fig. \ref{fig:topology}, our evaluation setup uses a combination of a simulation setup and an experimental testbed. We perform our evaluation in two steps. In the first step, we use ns3 \cite{ns3} to simulate the radio part of the 5G network where 320 mobile nodes generate UDP traffic and send it to the node \textbf{C} (see the Fig. \ref{fig:topology}). At node \textbf{C}, we collect all traffic traces destined for the slices in a \textit{pcap} file. This trace file collected from the simulation is used for all subsequent experiments that happen in the second step. In the second step which is performed on the testbed, we replay \textit{pcap} file at servers A1 and A2 using \textit{tcpreplay} \cite{tcpreplay} to perform DDoS attack on the slices. The two DDoS attack nodes A1 and A2 are used for a high bandwidth DDoS attack (at approximately 200Mb/s).

The reason for using the two-step approach is that we faced some challenges with scaling the experiments. The combination of a large amount of traffic generated by UEs and the emulated link between PGW and the rest of the network (slices), it became infeasible to perform DDoS using this setup. There was too much overhead in terms of processing requirement and network traffic. Therefore, we collected traffic traces at node \textbf{C} and used \textit{tcpreplay} at A1 and A2 to perform DDoS attack.

For 5G-specific components, we used ns3 \cite{ns3} with mmWave module~\cite{8344116} to simulate the 5G RAN network and used emulated link from Packet Data Gateway (PGW) to connect the slices. We were able to implement Network Slicing Selection Function (NSSF) that can select the slices based on the Network Slice Selection Assistance Information (NSSAI) received from the UEs or assign NSSAI to the UEs if it is unknown. In our ns3 implementation, if a UE does not know the NSSAI, the NSSF will assign the NSSAI, and it will redirect traffic to the correct slice (UEs-MME-NSSF-PGW-C). After the NSSAI assignment, the UEs will bypass the NSSF and directly go the destination slice (UEs-MME-PGW-C).

\subsection{Test Application}
To test the impact of the different attacks, we measure certain performance parameters for a test application that we implemented on the testbed. We implemented a simple authentication protocol. In our implementation, we assume that the client (Green node in Fig. \ref{fig:topology})  already has an NSSAI and it directly sends traffic to a specific slice.  In our experiments, all slices can receive Transmission Control Protocol (TCP) and User Datagram Protocol (UDP) traffic. We note that we are using SEAF, AUSF, and HSS to describe the basic function of our authentication mechanism. They are not complete implementation of 5G functions. Our HSS database has 1000 records, and it uses a binary search algorithm.

\begin{figure}[ht]
	\centering
	\includegraphics[width=8cm,keepaspectratio=true]{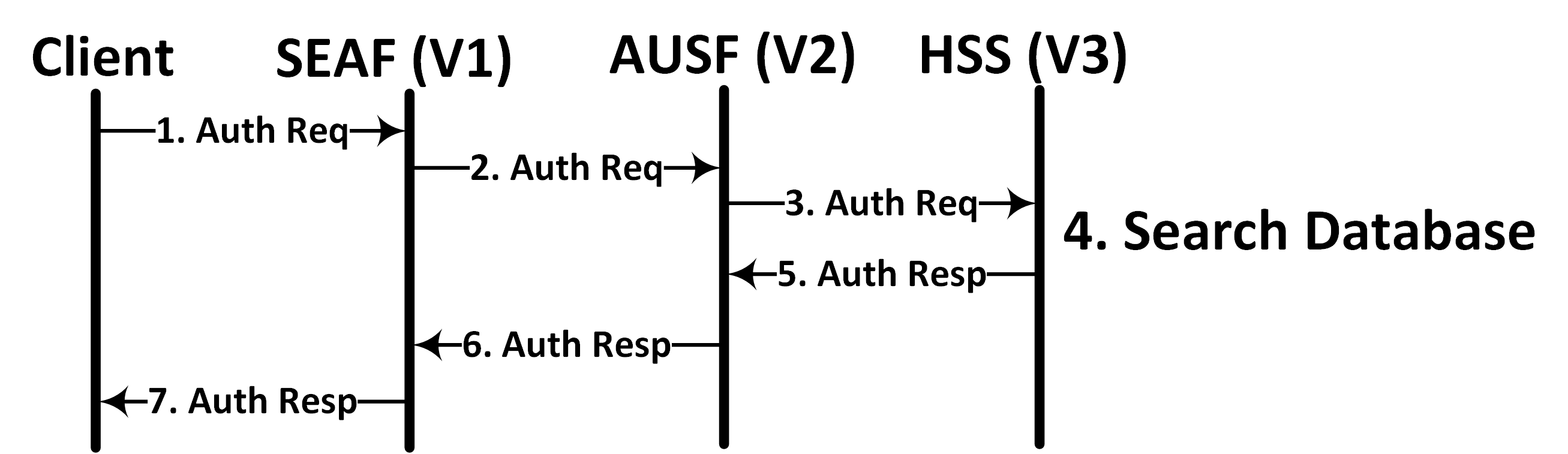}
	\caption{Flow diagram of the authentication procedure used in the experiments (V1, V2 and V3 represents virtual network functions)}
	\label{fig:flowdia}
\end{figure}

The messaging sequence of the authentication procedure is listed here and is also shown in Fig. \ref{fig:flowdia}.
\begin{enumerate}
	\item A client sends an authentication request to the slice SEAF.
	\item SEAF performs pre-processing and forwards it to the AUSF of the slice.
	\item AUSF forwards the request to the HSS and waits for the response.
	\item The HSS performs a database search
	\item HSS sends the result of the search to the AUSF
	\item AUSF sends success or failure message to the SEAF on the basis of information received from the HSS
	\item SEAF forwards the information to the client
\end{enumerate}

\subsection{Evaluation Methodology}
\label{sec:evalmethod}

To evaluate the performance impact of DDoS attack on slices, we collected response time and round-trip time (RTT) between the client and the slice and the average slice bandwidth available to the client. We used Ping between the client and slice S1 (SEAF) to calculate the RTT. To calculate the average available bandwidth for the entire slice, we measured bandwidth between client and slice S1 (SEAF) and between the slice S1's components using \emph{iperf} \cite{iperf}. The slice \emph{\textbf{S1}} is providing authentication service to a client and the slice \textbf{\emph{S4}} is under DDoS attack.

We empirically calculated the values for $T_{ef,init}$  and $\delta$ (eq. \ref{eq12a}) by measuring RTT between client and slice S1's components and between different components of the slice S1. We measure RTT between client and slice S1 as well as between different components of slice S1 when there is almost no traffic in the network and take an average of all collected data to estimate $T_{ef},init$. To estimate the value of $\delta$, we increase utilization of all available links for slice S1 to approximately 100\% and once again calculate RTT. We take an average of all the collected data, and it gives us a value of $\delta$.
The rest of the parameters used in the experiment are listed in Table \ref{tab:exparamter}.
\begin{table}[htbp]
	\centering
	\captionof{table}{Experiment parameters} \label{tab:exparamter}
	\begin{tabular}{r||l}\toprule[1.5pt]
		
		\bf Parameter		 & \bf Value 	\\\midrule
		\rowcolor{Gray}
		$N_p$				 &  3			\\
		$\sigma_k^{max}$	 &	10.0 GHz	\\
		\rowcolor{Gray}
		$K_{rel}$ 			 &	1-3		     \\
		$N_v$				 &	3			  \\
		\rowcolor{Gray}
		$R^{ij}$             &  5-12 Mbps	   \\
		$R^i$                &  0.4-1.4 GHz	\\
		\rowcolor{Gray}
		$\Delta^i$           &  0.1-1.0 ms	      \\
		$\Delta_k$           &  0.5 ms	        \\
		\rowcolor{Gray}
		$\delta$             &  16.85 ms	\\
		$T_{ef},init$        &  0.15 ms	\\
		\rowcolor{Gray}
		Total Slice Requests & 	12 		\\
		\bottomrule[1.25pt]
		\end {tabular}\par
		\bigskip
		\squeezeup
	\end{table}
	\section{Results and Discussion}
	\label{sec:resutls}
	\vspace{-0.1mm}
	In this section, we discuss our results. We devised two attack scenarios to realize the threat we described in section~\ref{sec:attacks}. In both scenarios, the client is sending one authentication request per sec to the slice S1 (SEAF), and the majority of the network traffic is directed towards either S1 or S4. To get a baseline (i.e., without isolation) results for comparison, we used a greedy algorithm. In the greedy algorithm, the slices are allocated on First Come First Serve basis. Once the PS1 is fully utilized, we start allocating slices to the next server and so on. In the figures, \emph{NoIsolation} represents the results of greedy allocation. Results for the different levels of intra-slice isolation are represented by $K_{rel}=1, K_{rel}=2,$ and $K_{rel}=3$.
	
	Due to the way our testbed is designed (with three physical servers), coincidentally, our optimization solution allocated slices in such a way that at intra-slice isolation level $K_{rel}=3$, the two slices we are concerned with in the evaluation (S1 and S4) were allocated entirely on two different servers which resulted in complete inter-slice isolation. Therefore, in all the presented results, $K_{rel}=3$ is also providing inter-slice isolation\footnote{We do have results (not presented here) obtained by setting $\gamma^c=1$ or $\gamma^d=1$. The difference in the results is approximately 1-2\% compared to ones presented here for $K_{rel}=3$. That could be due to the randomness of the network/CPU or other parameters. The results included in this paper are obtained with $\gamma^c=0$ or $\gamma^d=0$} (i.e., similar to setting $\gamma^c=1$ or $\gamma^d=1$). We note again that this coincidence is due to the way the testbed is set up and may not be repeated in other experimental settings.
	
	For each value of $K_{rel}$, we repeat every experiment three times. For every new value of $K_{rel}$, a new optimization solution is calculated, and slices are reallocated accordingly before the corresponding experiments are run. In our previous work \cite{DBLP:journals/corr/abs-1802-04655}, we discussed the overall CPU and bandwidth utilization of the system, and average solver runtime for different levels of intra-slice isolation.
	\subsection{DDoS flooding attack}
	
\begin{figure}[t!]
		\centering
		\begin{tikzpicture}[baseline]
		\begin{axis}[
		ymin=0,  ymax=300,
		grid=major,
		ylabel={Response Time (ms)},
		xlabel={Authentication Request \#},
		legend style={font=\scriptsize,at={(0.5,-0.28)},
			anchor=north,legend columns=-1},
		xtick distance=25,
		ytick distance=50,
		xmin=0,  xmax=150,
		]
		\addplot table[x=Request,y=DDoS] {FloodResponse.txt};
		\addplot table[x=Request,y=Krel-1] {FloodResponse.txt};
		\addplot table[x=Request,y=Krel-2] {FloodResponse.txt};
		\addplot table[x=Request,y=Krel-3] {FloodResponse.txt};
		\legend{NoIsolation,$K_{rel}=1$,$K_{rel}=2$,$K_{rel}=3$}
		\end{axis}
		\end{tikzpicture}
		\caption{Response Time for flooding DDoS attack scenario. Attack traffic starts at approximately authentication request\#30 and stops at approximately authentication request\#130}
		\label{fig:floodResponse}
	\end{figure}
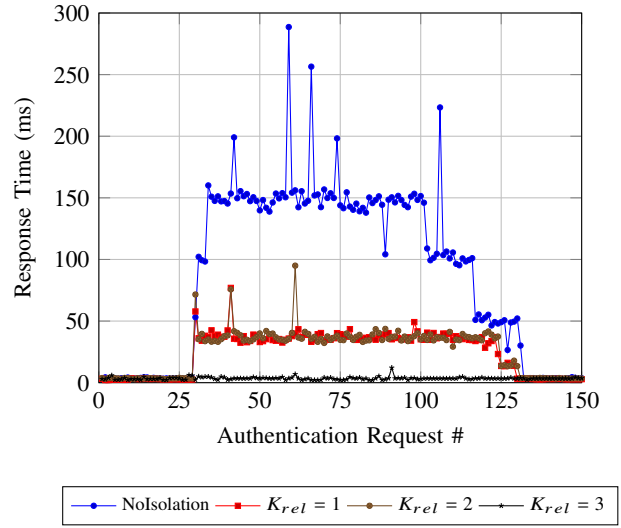

	In evaluating the DDoS flooding attack scenario \cite{8301700}. The target was slice S4 (attack details are in section~\ref{sec:FloodDDoS}).  Our client node is sending authentication requests to slice S1, and we use ns3 UDP traffic traces to launch a DDoS attack from A1 and A2 on slice S4. In this experiment, CPU is not a bottleneck.  Fig.~\ref{fig:floodResponse} shows the response time at the client. We can see a very high response time for NoIsolation and lowest for $K_{rel}=3$ (which also represents inter-slice isolation). When there is no isolation present between slices, all the components of different slices are sharing the same physical link. The Linux Traffic Controller \emph{tc} configures the kernel packet scheduler for all the virtual network function (VNF) hosted on the same machine and in this case a total of six components for both slices (for simplicity, we can ignore the other slices for our discussion). When slice S4 is under DDoS, we can see a significant impact on the response time of the slice S1's client. However, the effect is substantially reduced when $K_{rel}=1$ and $K_{rel}=2$.
	The main reason is that a packet spends less time waiting in the kernel queue for each VNF as compared to the NoIsolation case. For instance, in the NoIsolation scenario, there will be a minimum of six queues at the kernel (one for each VNF). Whereas, when $K_{rel}=1$ and $K_{rel}=2$, the minimum number of queues will be reduced from six to two and three, respectively. Another factor that improved the response time for the client is that there are multiple physical links available for the slice S1 to communicate with each other and the client. We do not see a trade-off (in terms of cost) of using $K_{rel}=1$ and $K_{rel}=2$ because the scale of the network is small and the solution for both values of  $K_{rel}$  is not the worst-case scenario (see section \ref{sec:worstC}). In previous work \cite{DBLP:journals/corr/abs-1802-04655}, we simulated a larger network where the cost of different values of $K_{rel}$ was more evident. In this paper, since we are using an actual testbed with a smaller scale, the cost is not evident. We expect it to show in a larger scale testbed. When $K_{rel}=3$, both slices are allocated on a separate physical server, and the traffic is completely isolated from the interference of other slices (inter-slice isolation), which provides the best results.
	
	\begin{figure}[t!]
		\centering
		\begin{tikzpicture}[baseline]
		\begin{axis}[
		grid=major,
		ymode=log,
		ylabel={Round-Trip Time (ms)},
		xlabel={Ping Request \#},
		legend style={font=\scriptsize,at={(0.5,-0.28)},
			anchor=north,legend columns=-1},
		xtick distance=25,
		ytick distance=10,
		xmin=0,  xmax=150,
		]
		\addplot table[x=Request,y=DDoS] {FloodRTT.txt};
		\addplot table[x=Request,y=Krel-1] {FloodRTT.txt};
		\addplot table[x=Request,y=Krel-2] {FloodRTT.txt};
		\addplot table[x=Request,y=Krel-3] {FloodRTT.txt};
		\legend{NoIsolation,$K_{rel}=1$,$K_{rel}=2$,$K_{rel}=3$}
		\end{axis}
		\end{tikzpicture}
		\caption{Round-Trip Time for flooding DDoS attack scenario. Attack traffic starts at approximately Ping request\#30 and stops at approximately Ping request\#130}
		\label{fig:floodRTT}
		\vspace{-4.0mm}
	\end{figure}
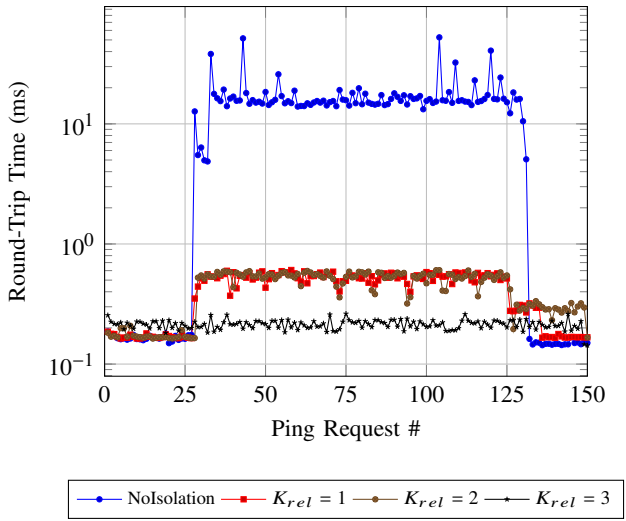
	We calculated RTT from the client to slice S1 (SEAF) as shown in Fig. \ref{fig:floodRTT}. We can see a significant advantage of our isolation solution at all level of $K_{rel}$ as compared to NoIsolation. However, the RTT is not as high as the response time (i.e, Fig. \ref{fig:floodResponse}). The RTT is calculated only from the client to the S1 (SEAF), but in the response time, an authentication request passes through all three nodes that could add significant delay (i.e., see authentication message flow in Fig. \ref{fig:flowdia}).
	
	\begin{figure}[t!]
		\centering
		\begin{tikzpicture}[baseline]
		\begin{axis}[
		ymin=0,  ymax=12,
		grid=major,
		ylabel={Average Available Bandwidth (Mbps)},
		xlabel={Elapsed Time (s)},
		legend style={font=\scriptsize,at={(0.5,-0.28)},
			anchor=north,legend columns=-1},
		xtick distance=25,
		ytick distance=2,
		xmin=0,  xmax=150,
		]
		\addplot table[x=Time,y=DDoS] {FloodBW.txt};
		\addplot table[x=Time,y=Krel-1] {FloodBW.txt};
		\addplot table[x=Time,y=Krel-2] {FloodBW.txt};
		\addplot table[x=Time,y=Krel-3] {FloodBW.txt};
		\legend{NoIsolation,$K_{rel}=1$,$K_{rel}=2$,$K_{rel}=3$}
		\end{axis}
		\end{tikzpicture}
		\caption{Average available bandwidth for the slice in flooding DDoS attack scenario. Attack traffic starts at approximately 30s and stops at approximately 130s}
		\label{fig:floodBW}
	\end{figure}
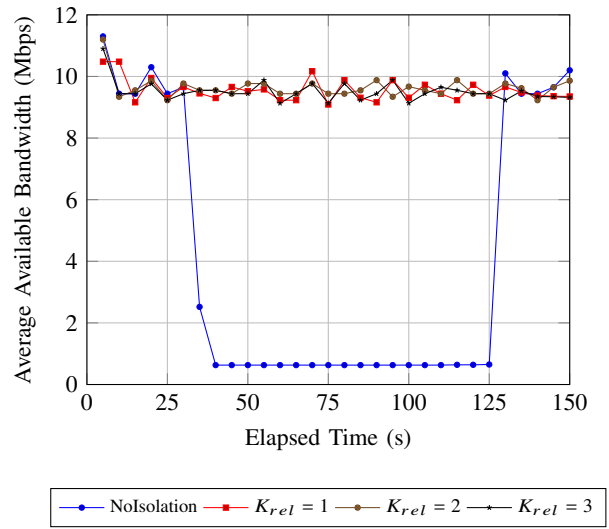
	Fig. \ref{fig:floodBW} shows the average available bandwidth in the entire slice S1. We measured bandwidth using \emph{iperf} between client and S1 (SEAF) and between slice S1's virtual network functions (we only plot average bandwidth). The slice S1 is allocated 10Mb/s bandwidth. As we can see, bandwidth for slice S1 drops to less then 1Mb/s for NoIsolation and these results also provide some insight for such high response time (Fig. \ref{fig:floodResponse}) and RTT (Fig. \ref{fig:floodRTT}). There is significantly less bandwidth available in the entire slice S1. However, our isolation solution guarantees a high percentage of the allocated bandwidth. The high bandwidth utilization at the beginning is due to initial burst traffic.
	
	\subsection{Slice-initiated attack}
	
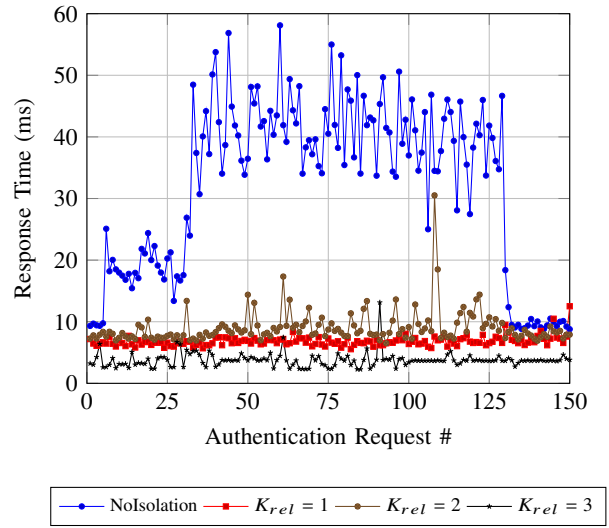
\begin{figure}[t!]
		\vspace{-2.5mm}
		\centering
		\begin{tikzpicture}[baseline]
		\begin{axis}[
		ymin=0,  ymax=60,
		grid=major,
		ylabel={Response Time (ms)},
		xlabel={Authentication Request \#},
		legend style={font=\scriptsize,at={(0.5,-0.28)},
			anchor=north,legend columns=-1},
		xtick distance=25,
		ytick distance=10,
		xmin=0,  xmax=150,
		]
		\addplot table[x=Request,y=DDoS] {CPUResponse.txt};
		\addplot table[x=Request,y=Krel-1] {CPUResponse.txt};
		\addplot table[x=Request,y=Krel-2] {CPUResponse.txt};
		\addplot table[x=Request,y=Krel-3] {CPUResponse.txt};
		\legend{NoIsolation,$K_{rel}=1$,$K_{rel}=2$,$K_{rel}=3$}
		\end{axis}
		\end{tikzpicture}
		\caption{Response Time for slice-initiated attack scenario. Attack traffic starts at approximately authentication request\#30 and stops at approximately authentication request\#130}
		\label{fig:CPUResponse}
	\end{figure}

\begin{figure}[t!]
		\centering
		\begin{tikzpicture}[baseline]
		\begin{axis}[
		ymin=0,  ymax=10,
		grid=major,
		ylabel={Round-Trip Time (ms)},
		xlabel={Ping Request \#},
		legend style={font=\scriptsize,at={(0.5,-0.28)},
			anchor=north,legend columns=-1},
		xtick distance=25,
		ytick distance=2,
		xmin=0,  xmax=150,
		]
		\addplot table[x=Request,y=DDoS] {CPURTT.txt};
		\addplot table[x=Request,y=Krel-1] {CPURTT.txt};
		\addplot table[x=Request,y=Krel-2] {CPURTT.txt};
		\addplot table[x=Request,y=Krel-3] {CPURTT.txt};
		\legend{NoIsolation,$K_{rel}=1$,$K_{rel}=2$,$K_{rel}=3$}
		\end{axis}
		\end{tikzpicture}
		\caption{Round-Trip Time for slice-initiated attack scenario. Attack traffic starts at approximately Ping request\#30 and stops at approximately Ping request\#130}
		\label{fig:CPURTT}
	\end{figure}
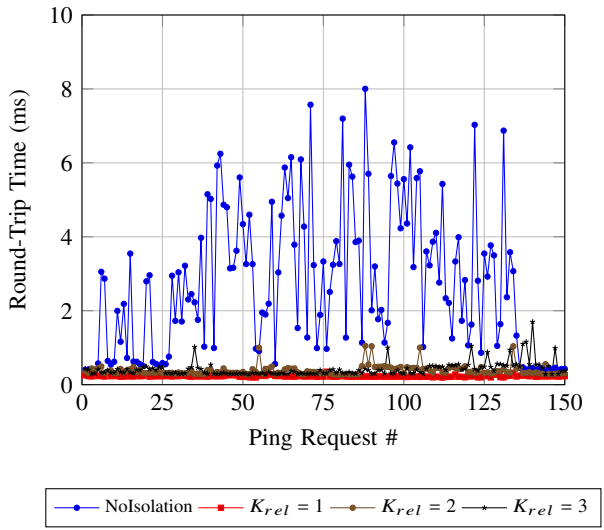

	In the slice-initiated attack scenario (see section~\ref{sec:sliceinitidated} for more details), we overload the CPU of slice S4 using a simple Python script and measure the impact on slice S1. There is no flooding DDoS attack happening in this scenario. The impact of CPU bottleneck depends on various factors, e.g., applications running on the host system and within containers (VMs), the host operating system and virtualization platform. In our experiments, we tried to keep the impact of the factors mentioned above to a minimum\footnote{We do not launch any additional applications between different experiments as well as at the beginning of each experiment, we measure CPU and RAM utilization to confirm that they are at a similar level.} and measure the response time and RTT for the client as shown in Fig. \ref{fig:CPUResponse} and Fig. \ref{fig:CPURTT}, respectively. As shown, the response time is relatively high for the client when there is no isolation (NoIsolation). Whereas, our isolation solution for all levels of $K_{rel}$ shows a minimal impact of sustained high CPU utilization. The minor variation between response times for different levels of $K_{rel}$ could be due to the factors we mentioned earlier.
	
	For RTT, we see much more variation when there is NoIsolation. The reason could be that when a Ping request arrives at the destination VNF, sometimes the CPU might be less loaded, and it can process the request sooner, and other times it might be more loaded and respond more slowly. However, CPU bottleneck has a minimal impact for all levels of $K_{rel}$. The impact of slice-initiated attacks could have been worse if other hosted slices were using some CPU intensive applications.
	\subsection{Current Limitations}
	\label{sec:worstC}
	
	{\bf Worst-case scenario allocation: } In our current solution, a worst-case scenario would be, if a significant number of VNFs from different slices that require either high bandwidth or perform CPU intensive tasks (could be a combination of both) are allocated on the same physical host(s). The allocation scheme do not take into consideration what type of VNFs are previously allocated on any specific host(s). So if they were of the resource-intensive type, and they happen to reach their maximum load at similar times, a DDoS attack on a slice at that time could pose significant mitigation challenges.
	
	\noindent {\bf Other limitations: } The allocation scheme is static as it does not re-allocate previously allocated slices to make more efficient use of available resources (or newly added resources). The migration of VNF(s) poses another challenge in the current solution. There is no mechanism in place for migrating a VNF(s). For VNF(s) migration, in some cases, it might be beneficial to migrate the under attack VNF(s) rather than other hosted VNF(s), or in other cases, it might be better to migrate slice components that are not under attack.
	Lastly, our allocation scheme does not consider the processing delay added by the intermediate nodes (e.g., router, switches, etc.).
	\section{Conclusion}
	\label{sec:con}
	In our solution, we used slice isolation to reduce the impact of DDoS attacks on a simple network service (slice authentication). We evaluated our solution using a combination of simulation and an experimental testbed for flooding DDoS attacks and slice-initiated attacks. In both attack scenarios, complete slice isolation provides the best mitigation. Whereas, at different intra-slice isolation levels $K_{rel}=1$ and $K_{rel}=2$, the impact of DDoS on the response time and round-trip time is significantly reduced as compared to the no isolation case (NoIsolation). During the DDoS attack scenario, the average bandwidth available to the client is a fraction of the requested bandwidth, when there is no isolation. However, the impact of DDoS on average available bandwidth with our solution enabled is minimal.
	
	We note that inter-slice isolation provides strong resource isolation. However, it can reduce the efficiency of resource utilization. The intra-slice isolation provides better control over the trade-off between security, availability and resource utilization.
	
	Currently, it is very challenging to do large scale 5G testing. We believe that our results, while obtained through a small testbed, have significant value in this new research area because they are obtained through actual experiments (rather than pure theoretical or pure simulation evaluation). We will continue to look for ways to scale up the testing experiments to see if the results will be the same for larger testbeds.
	
	\section*{Acknowledgment}
	This work was supported by the Natural Sciences and Engineering Research Council of Canada (NSERC) through the NSERC Discovery Grant program. We thank the anonymous reviewers for their comments and suggestions.
\bibliographystyle{IEEEtran}
\bibliography{5Gbib}
\end{document}